\documentclass[twocolumn,aps,prl,superscriptaddress,showpacs
]{revtex4}
\usepackage{bm,amsmath,amsfonts,amssymb,
graphicx,color}

\newcommand{\Eq}[1]{Eq.~\eqref{#1}}
\newcommand{\eq}[1]{\eqref{#1}}
\newcommand{\Fig}[1]{Fig.~\ref{#1}}

\newcommand{\beq}{\begin{equation}}
\newcommand{\eeq}{\end{equation}}
\newcommand{\beqa}{\begin{eqnarray}}
\newcommand{\eeqa}{\end{eqnarray}}
\newcommand{\Beqa}{\begin{eqnarray*}}
\newcommand{\Eeqa}{\end{eqnarray*}}

\newcommand{\sgn}{\text{sgn}}

\def\Xint#1{\mathchoice
   {\XXint\displaystyle\textstyle{#1}}%
   {\XXint\textstyle\scriptstyle{#1}}%
   {\XXint\scriptstyle\scriptscriptstyle{#1}}%
   {\XXint\scriptscriptstyle\scriptscriptstyle{#1}}%
   \!\int}
\def\XXint#1#2#3{{\setbox0=\hbox{$#1{#2#3}{\int}$}
     \vcenter{\hbox{$#2#3$}}\kern-.5\wd0}}
\newcommand{\pvint}{\Xint-}

\begin{document}

\title{Low-energy excitations of a one-dimensional Bose gas with weak contact repulsion}

\author{M. Pustilnik}
\affiliation{School of Physics, Georgia Institute of Technology, Atlanta, GA 30332, USA}
\author{K. A. Matveev}
\affiliation{Materials Science Division, Argonne National Laboratory, Argonne, IL 60439, USA}

\begin{abstract}
We study elementary excitations of a system of one-dimensional bosons with weak contact repulsion. We show that the Gross-Pitaevskii regime, in which the excitations are the  well-known Bogoliubov quasiparticles and dark solitons, does not extend to the low-energy limit. Instead, the spectra of both excitations have finite curvatures at zero momentum, in agreement with the phenomenological picture of fermionic quasiparticles. We describe analytically the crossover between the Gross-Pitaevskii and the low-energy regimes, and discuss implications of our results for the behavior of the dynamic structure factor. 
\end{abstract}



\maketitle

Recent developments in the physics of ultracold atomic gases~\cite{BDZ, Davidson} renewed the interest in fundamental properties of interacting bosons. This subject has a long history~\cite{Davidson,PS}. The first quantitative theory of elementary excitations was developed by Bogoliubov~\cite{Bogoliubov}, who found that quasiparticles in a weakly interacting Bose gas have linear spectrum at low momenta. This prediction was tested by Lieb and Liniger~\cite{LiebLiniger,Lieb} on the integrable model of one-dimensional bosons with contact repulsion. Surprisingly, the exact solution~\cite{Lieb} demonstrated the existence of not one, but two types of excitations. One of these excitations (type I in Lieb's classification) closely resembles~\cite{Lieb} the conventional Bogoliubov quasiparticle. The second (type II) excitation was subsequently identified~\cite{KMF,IT} with the dark soliton solution~\cite{soliton} of the time-dependent mean field equation by Gross and Pitaevskii~\cite{GP}.

Both excitation branches are acoustic with the same velocity $v$, but the leading nonlinear corrections to their spectra have opposite signs~\cite{Lieb,KMF}:
\begin{subequations}
\label{1}
\beqa
\varepsilon^+_p &=& vp + \frac{p^3}{8m^2 v},
\label{1a}\\
\varepsilon^-_p &=& vp - \frac{2}{5}\left(\frac{3}{4}\right)^{5/3}\!\!\frac{v p^{5/3}}{(\hbar n)^{2/3}}. 
\label{1b}
\eeqa
\end{subequations}
Here the superscript $+\,(-)$ denotes the type I (type II) excitations, $m$ is the mass of the constituent particles, and $n$ is their density. The Gross-Pitaevskii approach leading to \Eq{1} is applicable at weak repulsive interactions~\cite{PS}.

An alternative to the mean field approximation is the hydrodynamic approach~\cite{Popov,Haldane}. In this theory the excitations of a one-dimensional quantum liquid are waves of density (phonons). In the harmonic approximation the phonon spectrum coincides with that of the Bogoliubov quasiparticle $\varepsilon^+_p$, and at $p\ll mv$ is given by \Eq{1a}. However, this result may not hold in the limit $p\to 0$ because among various anharmonic terms neglected in this approximation there are irrelevant perturbations of scaling dimension $3$. At small $p$ they scale as $p^2$, and their effect may exceed the $p^3$ correction in \Eq{1a}. To account for these perturbations it is convenient to apply the well-known mapping~\cite{Mattis} between one-dimensional bosons and fermions, in which the phonons are represented by superpositions of particle-hole pairs. The resulting effective Hamiltonian describes weakly interacting fermions with quadratic spectrum~\cite{Rozhkov}. Thus, there are two excitation branches, particles and holes, with dispersion relations
\beq
\varepsilon^{\pm}_p = vp \pm \frac{p^2\,}{2m_*}.
\label{2}
\eeq
The phenomenological approach~\cite{Rozhkov,ISG} leading to \Eq{2} is applicable to any single component one-dimensional quantum fluid, regardless of the statistics of the constituent particles. For weakly interacting bosons the effective mass $m_*$ in \Eq{2} is given by~\cite{Pereira,ISG} 
\beq
\frac{m\,}{\,m_*}  = \frac{3}{4} \sqrt{\frac{\,mv}{\pi \hbar n}}.
\label{3}
\eeq
Interestingly, for both excitation branches the nonlinear corrections in Eqs.~\eq{1} and \eq{2} are of the same order of magnitude at $p\sim p_*$, where
\beq
p_* = \frac{4m^2v\,}{\,\,m_*}.
\label{4}
\eeq  
Therefore, it is natural to conjecture~\cite{ISG,AP} that Eqs.~\eq{1} and \eq{2} refer to the same excitations at $p\gg p_*$ and $p\ll p_*$, respectively. 

In this paper we derive analytic expressions for the spectra of elementary excitations of one-dimensional bosons with weak contact repulsion at momenta $p \ll mv$. This range includes the crossover scale $p_*$. On the two sides of the crossover, our expressions match Eqs.~\eq{1} and \eq{2}, thereby verifying the above conjecture. The spectra $\varepsilon_p^\pm$ determine the position and the nature of power-law singularities in the dynamic structure factor~\cite{khodas,shashi,IG,ISG}. This quantity can be measured by Bragg spectroscopy in ultracold atomic gases~\cite{BDZ,Davidson,Bragg}. 

We consider the Lieb-Liniger model~\cite{LiebLiniger,Lieb} 
\beq
H = \frac{\,\hbar^2}{2m\,}\left[
-\sum_{i=1}^N\frac{\,\partial^2}{\partial x_i^2}
+ c\sum_{i\neq j}\delta(x_i - x_j)
\right],
\label{5}
\eeq
describing $N$ bosons in a system of size $L$ with periodic boundary conditions. We are interested in the thermodynamic limit $N,L\to\infty$ taken at fixed density $n = N/L$. Weak repulsion corresponds to $\gamma = c/n\ll 1$. In this regime the velocity of the elementary excitations is given by $v = (\hbar n/m)\gamma^{1/2}$~\cite{PS,Lieb}.

The model \eq{5} is integrable by Bethe ansatz~\cite{LiebLiniger,Lieb,Yang}. In this technique the exact many-body eigenstates are characterized by rapidities, which are similar to the wave numbers of noninteracting fermions. Lieb's type I and type II excitations are analogous to the particle and hole excitations of the corresponding Fermi gas~\cite{Lieb}. Their momentum $p$ and energy $\varepsilon$ are given parametrically by
\beq
p = 2\pi\hbar\left|\int_{k_0}^{k}\!dk'\rho(k')\right|,
\quad
\varepsilon = \left|\int_{k_0}^k\!dk'\sigma(k')\right|,
\label{6}
\eeq
where $|k| > k_0$ $\bigl(|k| < k_0\bigr)$ for the type I (type II) excitations. In \Eq{6} $\rho(k)$ is the density of rapidities in the ground state and $\sigma(k)$ is the derivative of the energy function introduced in~Ref.~\cite{Yang}. The functions $\rho(k)$ and $\sigma(k)$ satisfy the integral equations~\cite{LiebLiniger,Lieb,Yang}
\begin{subequations}
\label{7}
\beqa
\rho(k) - \frac{c}{\pi}\!\int_{-k_0}^{k_0}\!dk' \frac{\rho(k')}{(k-k')^2 + c^2} 
&=& \frac{1}{2\pi},
\label{7a}
\\
\sigma(k) - \frac{c}{\pi}\!\int_{-k_0}^{k_0}\!dk' \frac{\sigma(k')}{(k-k')^2 + c^2} 
&=&  \frac{\hbar^2 k}{m}\,.
\label{7b}
\eeqa
\end{subequations}
The value of the  Fermi rapidity $k_0$ in Eqs.~\eq{6} and \eq{7} is set by the condition $\int_{-k_0}^{k_0}\!dk\,\rho(k) = n$.

At the level of the Bethe ansatz equations \eq{7}, the Gross-Pitaevskii approximation amounts to finding asymptotic solutions of Eqs.~\eq{7a} and \eq{7b} at $c$ approaching zero~\cite{LiebLiniger,Lieb,KMF,IT}. Such solutions have the form
\begin{subequations}
\label{8}
\beqa
\rho(k) &=& \sgn(k)\,\frac{1}{2\pi\,}\frac{d}{dk}(k^2 - k_0^2)^{1/2},
\label{8a}
\\
\sigma(k) &=& \sgn(k)\,\frac{\,\hbar^2}{6m\,} \frac{d^2}{dk^2} (k^2 - k_0^2)^{3/2}
\label{8b}
\eeqa
\end{subequations}
at $|k| > k_0$ and
\begin{subequations}
\label{9}
\beqa
\rho(k) &=& \frac{1}{2\pi c}(k_0^2 - k^2)^{1/2}, 
\label{9a}
\\
\sigma(k) &=& -\,\frac{\,\hbar^2}{6mc} \frac{d}{dk}(k_0^2 - k^2)^{3/2}
\label{9b}
\eeqa
\end{subequations}
at $|k| < k_0$, with the Fermi rapidity $k_0 = 2n \gamma^{1/2}$. 

Substitution of Eqs.~\eq{8} and \eq{9} into \Eq{6} reproduces the spectra of the elementary excitations obtained in Ref.~\cite{KMF} from the exact solution of the Gross-Pitaevskii equation. In order to obtain the leading nonlinear corrections [see \Eq{1}] it is sufficient to expand Eqs.~\eq{8} and \eq{9} in small $q = k-k_0$. For example, at $0 < q \ll k_0$ \Eq{8} yields
\begin{subequations}
\label{10}
\beqa
\rho(q) 
&=& \frac{1}{2\pi}\sqrt{\frac{k_0}{2\,}}
\left(
q^{-1/2} + \frac{3}{4k_0}\, q^{1/2}
\right),
\quad
\label{10a}
\\
\sigma(q) 
&=&
\,\hbar v_{}\sqrt{\frac{k_0}{2\,}}
\left(
q^{-1/2} + \frac{15\,}{\,4k_0}\, q^{1/2}
\right).
\label{10b}
\eeqa
\end{subequations}
Substitution of the expansions \eq{10} into \Eq{6} indeed results in \Eq{1a}. Similarly, expanding \Eq{9} in small $|q|$ at $0 < -\,q \ll k_0$ leads to \Eq{1b}. Note that the ratio $\sigma/\rho$ becomes $2\pi\hbar v$ in the limit $|q|\to 0$, which gives $\varepsilon_p^\pm = vp$ at $p\to 0$. Thus, in order to recover the nonlinear corrections in Eq.~\eq{1}, it is necessary to retain the subleading terms in the expansions \eq{10}. 

In addition to the condition $q\ll k_0$, the range of applicability of \Eq{10} is also restricted at small $q$. Indeed, expansions \eq{10} diverge in the limit $q\to 0$. The divergence originates in the fact that solutions \eq{8} and \eq{9} are valid only asymptotically at $c\to + \,0$. On the other hand, exact solutions of the Bethe ansatz equations \eq{7} must be analytic at all $k$, including $k = \pm\,k_0$. Therefore, the validity of the solutions \eq{8} and \eq{9} is limited to $|q| = |k - k_0|\gg c$. The restriction on $|q|$ translates to the restriction on the momentum. From \Eq{10a}, the density of rapidities at $|q|\sim c$ is estimated as $\rho\sim (k_0/c)^{1/2} \sim \gamma^{-1/4}$. Equation \eq{6} then shows that the corresponding momentum is of order $\hbar n\gamma^{3/4} \sim p_*$, where $p_*$ is given by \Eq{4}. Thus, the condition $|q|\gg c$ corresponds to $p\gg p_*$ in \Eq{1}.

The Bethe ansatz equations \eq{7} simplify considerably at $|q| = |k - k_0|\lesssim c\ll k_0$. In this regime the right-hand sides of Eqs.~\eq{7a} and \eq{7b} can be neglected, and the lower limits of the integrations can be extended to infinity~\cite{Popov_LL}. Both equations then assume the form
\beq
f(q) = \frac{c}{\pi}\!\int_{- \infty}^0\!dq'\frac{f(q')}{(q-q')^2 + c^2}.
\label{11}
\eeq
Unlike the original Bethe ansatz equations \eq{7}, this integral equation is of Wiener-Hopf type, and its solutions can be found analytically. Such analysis allows us to obtain the behavior of $\rho$ and $\sigma$ at $|q|\sim c$. The resulting expressions read
\begin{subequations}
\label{12}
\beqa
\rho(q) &=& \frac{1}{2\pi}\sqrt{\frac{k_0}{c\,}}
\left[\varphi_0(q/c) + \frac{3c\,}{\,8k_0}\,\varphi_1(q/c)\right] ,
\label{12a}
\\
\sigma(q) &=& \hbar v\,\sqrt{\frac{k_0}{c\,}}\,\left[\varphi_0(q/c) 
+ \frac{15 c\,}{\,8k_0}\,\varphi_1(q/c)\right].
\label{12b}
\eeqa
\end{subequations}
The dimensionless function $\varphi_0$ in \Eq{12} is defined as
\beq
\varphi_0(t) = 
\int_0^\infty\!\!\frac{dz\,}{2\pi z^{1/2}}
\sin(2\pi z)\Gamma(z)\,e^{-z(\ln z -1+ 2\pi t)} 
\label{13}
\eeq
at $t > 0$ and 
\beq
\varphi_0(t) =
\pvint_0^\infty\!\!\frac{dz\,}{2\pi z^{3/2}}\!
\left[1 - \frac{\pi e^{z(\ln z -1+ 2\pi t)}}{\tan(\pi z)\Gamma(z)}\right]
\label{14}
\eeq
at $t < 0$. Simple poles at integer $z$ in the integrand of \Eq{14} are to be treated as Cauchy principal values. 

The function $\varphi_0(t)$ decreases monotonically from $\varphi_0(t) = |2t|^{1/2}$ at $-\,t\gg 1$ to $\varphi_0(t) = (2t)^{-1/2}$ at $t\gg 1$. Accordingly, at $q\gg c$ the first terms in the right-hand sides in \Eq{12} coincide with the leading contributions in the expansions~\eq{10}. Similar to \Eq{10}, it is necessary to keep the subleading terms in \Eq{12} in order to obtain the nonlinear corrections to the excitation spectra. These terms are described by the dimensionless function 
\beq
\varphi_1(t) = \int_0^t dt' \varphi_0(t').
\label{15}
\eeq  

Substituting \Eq{12} into \Eq{6}, we obtain the spectra in the form
\beq
\varepsilon^{\pm}_p = vp + \frac{p_*^2\,}{2m_*}\,e^{_{}\pm}(p/p_*),
\label{16}
\eeq
where the dimensionless functions $e^{_{}\pm}(s)$ are defined parametrically by the equations
\beq
s = \frac{\sqrt{2\pi}}{\,3}\,\bigl|\varphi_1(\pm\,\tau)\bigr|,
\quad
e^{_{}\pm} = \frac{\,(2\pi)^{3/2}}{9\,\,} 
\!\int_0^{\pm\,\tau}\!\!dt\, \bigl|\varphi_1(t)\bigr|
\label{17}
\eeq
with $0<\tau <\infty$.
  
\begin{figure}[t]
\centering
\includegraphics[width=.999\columnwidth]{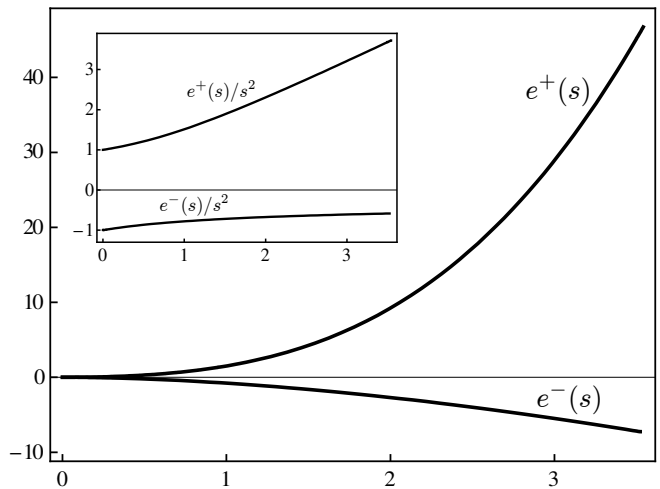}
\caption{
The crossover functions $e^\pm(s)$ [see Eqs.~\eq{15} and \eq{17}] describing the nonlinear in momentum corrections to the spectra of Lieb's type I and type II excitations. The inset illustrates that $e^\pm(s)\simeq \pm\,s^2$ at $s\to 0$ [see \Eq{19}].
}
\label{Fig1}
\end{figure}

Equations \eq{16} and \eq{17} represent the main result of this paper. They provide a complete analytic description of the spectra of elementary excitations of a weakly repulsive Bose gas in the entire range of momenta $p\ll mv$, which includes the crossover between the true low-energy regime at $p\ll p_*$ and the Gross-Pitaevskii regime at $p\gg p_*$. The crossover is described by the functions $e^{\pm}(s)$ plotted in \Fig{Fig1}. At $s\gg 1$ the crossover functions behave as
\begin{subequations}
\label{18}
\beqa
e^+(s) &=& s^3 + \frac{2}{3}s+\cdots,
\label{18a} \\
e^-(s) &=& - \,\frac{3}{5}\!\left(\frac{2\pi}{3}\right)^{2/3}\!
s^{5/3} + \frac{2}{9}s+\cdots.
\label{18b}
\eeqa
\end{subequations} 
Substituting the leading terms of these expansions into \Eq{16}, we recover \Eq{1}.
At $s\ll 1$ \Eq{17} yields
\beq
e^{_{}\pm}(s) = \pm\,s^2 + \frac{1}{3}s^3 +\cdots,
\label{19}
\eeq
and the spectra \eq{16} assume the form
\beq
\varepsilon^{\pm}_p = vp \pm \,\frac{p^2\,}{2m_*} \,+\,\frac{p^3}{24 m^2 v} + \cdots
\label{20}
\eeq
at $p\ll p_*$, in agreement with \Eq{2}.

As mentioned above, the spectra $\varepsilon^\pm_p$ reveal themselves in the behavior of the dynamic structure factor 
\beq
S(p,\varepsilon) = \int\!dx\,dt\,e^{i(px - \varepsilon t)/\hbar}\bigl\langle n(x,t)n(0,0)\bigr\rangle,
\label{21}
\eeq
where $n(x,t)$ is the density operator in the Heisenberg picture. At zero temperature, $S(p,\varepsilon)$ exhibits a non-analytic dependence on $\varepsilon$ at $\varepsilon = \varepsilon^{\pm}_p$~\cite{khodas,shashi,IG,ISG}. In particular, the type II excitation  $\varepsilon = \varepsilon^-_p$ serves as a threshold for the structure factor: $S(p,\varepsilon) = 0$ at $\varepsilon < \varepsilon^-_p$, reflecting the fact that $\varepsilon^-_p$ is the lowest possible energy for a state with momentum $p$. Close to the threshold, at $0\leq \varepsilon - \varepsilon^-_p\ll vp -  \varepsilon^-_p$, the dependence of the structure factor on $\varepsilon$ is governed by the power law~\cite{khodas,shashi,IG,ISG}
\beq
S(p,\varepsilon) \propto (\varepsilon - \varepsilon^-_p)^{\mu_p}. 
\label{22}
\eeq
The momentum-dependent exponent $\mu_p$ in \Eq{22} can be expressed via the spectrum $\varepsilon^-_p$~\cite{ISG,IG}. Using \Eq{20} and the relations obtained in~Ref. \cite{IG}, we find
\beq
\mu_p = \frac{p\,}{\,p_*},
\quad
p\ll p_*
\label{23}
\eeq
in the most interesting regime of small momenta. Note that this result cannot be obtained from \Eq{2}: The cubic in $p$ correction in \Eq{20}, found in this paper, makes the dominant contribution to the exponent at small $\gamma$.

Interestingly, the behavior described by \Eq{23} is typical of weakly interacting spinless fermions~\cite{PKKG}. In this case the linear dependence of the exponent on $p$ is a direct consequence of the Pauli principle. Thus, although our study did not address directly the statistics of the quasiparticles, \Eq{23} strongly supports the conjecture that at $p\ll p_*$ it is fermionic.   

To conclude, in this paper we studied the spectra of the elementary excitations of one-dimensional bosons with weak contact repulsion. The existence of the low-energy regime beyond the reach of the Gross-Pitaevskii equation, initially suggested on phenomenological grounds, has been demonstrated for a microscopic model. We described analytically the entire crossover between this regime and the Gross-Pitaevskii regime at higher energies. 
Our results can be tested by Bragg spectroscopy measurements of the dynamic structure factor. We found its behavior to be consistent with the fermionic nature of the low-energy excitations.  

\begin{acknowledgments}
This work was supported by the U.S. Department of Energy, Office of Science, Materials Sciences and Engineering Division. The authors are grateful to the Aspen Center for Physics (NSF Grant No. PHYS-1066293) for hospitality.
\end{acknowledgments}


\end{document}